\documentstyle[preprint,aps]{revtex}
\begin{document}
%\twocolumn[
%\hsize\textwidth\columnwidth\hsize\csname@twocolumnfalse\endcsname
%\draft

\title{On the c-axis optical reflectivity of layered cuprate
superconductors} 
\author{S.\ Das Sarma and E. H.\ Hwang}
\address{Center for Superconductivity Research,Department of Physics \\
	 University of Maryland, College Park, Maryland  20742-4111 }
\date{\today}
\maketitle

\begin{abstract}
Using a conventional BCS -- Fermi liquid model we calculate the
c-axis optical reflectivity of the layered high temperature cuprate
superconductors by obtaining the finite temperature 
dynamical dielectric function in a microscopic self-consistent gauge
invariant formalism. We get good semi-quantitative agreement with all
the existing experimental data by using the measured normal state $dc$
resistivities as the input parameters in obtaining the c-axis hopping
amplitude and the normal state level broadening in our microscopic
calculation. 

\end{abstract}
%\vspace{0.5in}
%]

\newpage

The c-axis charge dynamics of high-temperature layered cuprate
superconductors is unusual and interesting, and has therefore
attracted a great deal of well-deserved attention. In this Letter we
present a microscopic theoretical calculation for the c-axis optical
reflectivity of the cuprates, concentrating on the optimum doping
(normally corresponding to the highest critical temperature $T_c$)
single layer (eg. LSCO, Tl-2201) cuprate systems which have only one
CuO layer per unit cell. An impressive series\cite{one,two,three} of
c-axis optical reflectivity measurements by Uchida and co-workers on
single-crystal $La_{1-x}Sr_xCuO_4$ (LSCO) systems shows the following
salient features in the frequency dependence of the electronic
reflectivity: (1) For temperatures $T>T_c$, {\it i.e.}, in the normal
phases, the c-axis optical reflectivity is essentially structureless
down to 20 $cm^{-1}$ (the lowest frequency studied in
ref. \onlinecite{one}), and the data resemble that of an ionic
insulator; (2) in the superconducting phase, for $T<T_c$, the
reflectivity develops a well-defined sharp plasma edge with the
effective plasma frequency, $\omega_{pl}$ (defined as the frequency
where the real part of the dielectric function goes through a zero,
Re$\epsilon(\omega_{pl})=0$), being below the superconducting energy
gap $\Delta$ (in fact, $\omega_{pl} \sim 50 cm^{-1}$ in optimally
doped LSCO) -- thus the superconducting state exhibits a classic
``metallic'' optical reflectivity behavior in contrast to the normal
state ``insulating'' behavior; (3) the very low c-axis plasma
frequency (below the energy gap) associated with the observed
reflectivity plasma edge corresponds to a carrier mass substantially
larger than the c-axis optical mass obtained in standard band
structure calculations, {\it i.e.,} the experimental $\omega_{pl}$ is
much smaller, by more than an order of magnitude, than the
corresponding local-density-approximation (LDA) band structure calculated
value; (4) the superconducting state plasma edge shows a characteristic
temperature dependence, becoming weaker (and $\omega_{pl}(T)$ shifting
to lower frequency with increasing temperature) as $T$ approaches
$T_c$ from below; (5) the plasma edge shows a strong doping
dependence. It has been emphasized that these anomalies, in
particular, items (1)-(3) above, are very difficult to understand on
the basis of the standard Fermi liquid BCS-type generic theories of high
temperature superconductors.

In the theoretical work presented here we obtain the c-axis
reflectivity of the layered cuprate materials by calculating the
appropriate long wavelength dynamical dielectric function
$\epsilon(\omega)$ of the system.
In calculating the long wavelength frequency
dependent dielectric function we use the Nambu-Gorkov
formalism\cite{five} and carry out a fully gauge invariant
self-consistent finite temperature linear response calculation
\cite{six} including ``impurity scattering'' induced level broadening
effects in the theory\cite{seven}. Our theory, therefore, corresponds to the
one\cite{six} developed by Fertig and Das Sarma for anisotropic
layered superconducting systems except for three important
generalizations beyond ref. \onlinecite{six}: (1) we use a finite
temperature formalism in contrast to the $T=0$ case studied in
ref. \onlinecite{six}, (2) we use the d-wave order parameter symmetry
for the superconducting ground state in our response calculation
\cite{eight}  in contrast to the $s$-wave model used in
ref. \onlinecite{six} (this turns out to be unimportant in
understanding reflectivity measurements), and (3) we include level
broadening effects in our microscopic calculation in contrast to
ref. \onlinecite{six} which neglected disorder. We point out that the
$d$-wave symmetry makes negligible difference\cite{eight} with respect
to the $s$-wave symmetry in our long wavelength dielectric response
calculation (there are substantial differences\cite{eight} at large
wave vectors, which are not of any significance to optical reflectivity
measurements). Inclusion of level broadening in the dynamical
dielectric function is, however, crucial in understanding the c-axis
reflectivity measurements -- in particular, strong disorder scattering
in these systems, as reflected in their anomalously high c-axis $dc$
resistivities for $T>T_c$, gives rise to an overdamped plasmon in the
normal state, completely suppressing any normal state metallic plasma
edge for $T>T_c$. A plasma edge in the superconducting state
corresponding to the c-axis Anderson-Bogoliubov plasma
mode\cite{six,eight} is, however, preserved because single particle
level broadening does not affect the condensed carriers in the
superconducting state.

We assume\cite{six} a simple single tight binding band (of bandwidth
$2t_c$) along the c-axis and take the single-particle energy dispersion
to be $E(k,k_z) = \hbar^2 k^2/2m - t_c \cos (k_zd)$, where
$k\equiv|{\bf k}_{ab}|$ is the wave vector in the $ab$ plane (with $m$ as
the planar effective mass for charge dynamics in the $ab$ plane), $d$ is
the c-axis layer separation, and $k_z$  the wave vector along the
c-axis. Following refs. \onlinecite{six} and \onlinecite{eight} we
carry out a self-consistent (and gauge invariant) linear response
calculation for the dynamical dielectric function $\epsilon(\omega)$
treating the long range Coulomb interaction (in  the highly
anisotropic layered system) in the random phase approximation
(``bubble'' diagrams) and the BCS-like short-range pairing
interaction in the ladder approximation (which is a conserving
approximation because the BCS self-energy producing the
superconductivity is a Hartree-Fock single-loop self-energy). 
We include level broadening effects both in the single-particle
Green's function and in the vertex correction through the ladder
impurity diagrams\cite{seven}. Assuming the impurity scattering to be
short-ranged 
and isotropic, we can parameterize the disorder strength by a single
level broadening parameter ({\it i.e.,} an imaginary part of the
self-energy) $\gamma$ where $\tau = (2\gamma)^{-1}$, with $\hbar=1$
throughout, is the transport relaxation time for the c-axis normal
transport properties. Thus in principle, the normal state c-axis
resistivity $\rho_c$ of the system defined $\gamma$ within our
model. The calculated microscopic $\epsilon(\omega)$ therefore depends
on $\gamma$, $T$, $\Delta$ and $T_c$ (which are determined
\cite{six,eight} by the 
strength of the pairing interaction in our BCS model) as well as on
the effective single-particle parameters $m$ and $t_c$. The results
depend also on a number of obvious parameters such as the
two-dimensional carrier density $n_{2D}$ (which determines the
density of states), the layer separation $d$ in the c-direction, and
the effective background dielectric constant $\kappa$. Our philosophy
at this stage is to treat $\gamma$ and $t_c$, the two most contentious
parameters in our theory, as unknown parameters, and take the other
parameters from the best available experimental (and where
appropriate, from band structure) results. 
We note that the details of the pairing interaction or the impurity
potential do not affect our reflectivity results in any qualitative
way. 

In Fig. \ref{fig1} we show our calculated c-axis optical
reflectivity to be compared directly with the experimental results of
Tamasaku {\it et. al.}\cite{one}. We have use parameters corresponding
to LSCO except for $t_c$ and $\gamma$ which are varied to produce
different curves in Fig. \ref{fig1}. Our calculated results look {\it
very similar} to the experimental observations (see, for example,
Fig. 1 in ref. \onlinecite{one}), and in fact our theory agrees with the
measurements {\it quantitatively} if we treat $t_c$ and $\gamma$ as
phenomenological parameters and choose $t_c \approx 0.1 \Delta -0.3 \Delta$
and $\gamma \geq \Delta$. Typically we find the plasma edge
$\omega_{pl} \approx 2t_c - 4t_c$ in the superconducting state, and
provided $\gamma \geq \Delta$ no plasma edge shows up in the normal
state with the normal state reflectivity remaining essentially
constant around 0.5 (exactly as in the experimental results of
ref. \onlinecite{one}) down to rather low frequencies ($\omega \leq
0.1 \Delta$). We predict that if the normal state reflectivity
measurement is pushed down to rather low frequencies ($\sim 10
cm^{-1}$ or below), one would see a rise in the reflectivity as it
would increase to unity to be consistent with there being a finite
(albeit very low) $dc$ conductivity along the c-axis.

It is clear from the results shown in Fig. \ref{fig1} that to get
agreement with the experimental measurements\cite{one,two,three}
within our model one must have a very low (high) value of the c-axis hopping
amplitude $t_c$ (level broadening parameter $\gamma$). In particular,
our theory gives $\omega_{pl} \sim 2 t_c$ in the superconducting
state, which constraints $t_c$ to be in the range of 20 - 30 $cm^{-1}$
in LSCO for the theory to be quantitatively consistent with the experimental
measurement in ref. \onlinecite{one} which finds a sharp reflectivity
plasma edge in the 30 - 60 $cm^{-1}$. The other striking experimental
anomaly\cite{one} of the nonexistence of any reflectivity plasma edge
in the normal state is ``explained'' in our theory by having a large
damping or broadening term $\gamma$ ($\geq \Delta$) which enters our
microscopic impurity scattering diagrams in the dynamical
polarizability function. We have so far treated $t_c$ and $\gamma$ as
phenomenological parameters, and the crucial issue therefore is to
justify (or, at least provide a rationale for) having such small (high)
values of $t_c$ ($\gamma$) in layered high temperature superconducting
systems. This, in fact, was the basis of the trenchant analysis of the
experimental results\cite{one,two,three} carried out by
Anderson\cite{four}, which led him to conclude\cite{four} that a
standard Fermi liquid-BCS type analysis (exactly of the type we carry
out in this work) is incapable of explaining the experimental c-axis
reflectivity data.

In this work we treat $t_c$ and $\gamma$ as phenomenological
parameters to be fixed by {\it normal state} $dc$ transport
properties of the systems. We use the {\it measured} values of
normal state $\rho_c$ and $\rho_{ab}$ at $T \geq T_c$ to fix $t_c$ and
$\gamma$ within the standard Fermi liquid picture. The hopping
amplitude $t_c$ is then given simply by the formula 
\begin{equation}
t_c = \left ( \frac{\pi n_{2D}}{2d^2 m^2} \frac{\rho_{ab}}{\rho_c}
\right )^{1/2}. 
\label{tc}
\end{equation}
The level broadening $\gamma$ is given simply by the c-axis $dc$
resistivity through the usual relaxation time approximation in the
tight binding limit:
\begin{equation}
\gamma = \frac{e^2n_{2D}dt_c\rho_c}{2} = \frac{e^2}{2m} \sqrt{
\pi n_{2D}^3 \rho_c \rho_{ab}/2}.
\label{gam}
\end{equation}
The simple Drude-Boltzmann-Fermi liquid formulae given by Eqs.
(\ref{tc}) and (\ref{gam}) immediately explain why the hopping
parameter $t_c$ (the level broadening parameter $\gamma$) is extremely
small (large) -- both of these anomalies arise from the anomalously
large c-axis resistivity, which is a universal feature for the layered
cuprates. We should emphasize two significant aspects of the
phenomenological analysis giving as the all-important parameters $t_c$
and $\gamma$ (from the normal state $dc$ resistivities of the system):
(1) standard LDA band structure calculations\cite{nine} give values of
$t_c$ which are one to two orders of magnitude larger than those given
by Eq. (\ref{tc}) for layered cuprates (while the two dimensional
electron density and the effective mass in the $ab$ plane seem to be
reasonably, {\it i.e.}, within a factor of two, well-described by LDA
band calculations\cite{nine}); (2) our Drude-Fermi liquid
phenomenological description characterized by Eqs. (\ref{tc}) and
(\ref{gam}) is completely independent of the actual normal state
transport mechanism along the c-axis and depends only on the crucial
observation that the normal state $dc$ transport along the c-axis is
``metallic'' around optimal doping (albeit with an anomalously large
imaginary part of the self-energy $\gamma$).

In Table 1 we summarize our results for a number of layered cuprates, 
giving the values of the parameters $t_c$ and $\gamma$ ( obtained from
experimental $\rho_c$ and $\rho_{ab}$ at $T \geq T_c$) used in our
calculations to obtain $\omega_{pl}$ at $T=0$, the long wavelength
c-axis Anderson-Bogoliubov plasmon frequency in the superconducting
phase.  In carrying out our microscopic calculation for the
``multilayer'' cuprates ({\it e.g.} YBCO, BISCO), which contain more
than one CuO layers per unit cell, we are concentrating here only on
the low frequency c-axis plasma properties which arise from the {\it
intercell} hopping, neglecting the mode mixing effects arising from
the {\it intracell} hopping between the CuO layers. This is
equivalent to making an effective single layer model for these
multilayer cuprates, which should suffice for the c-axis reflectivity
properties. From the available experimental results given in Table 1
it is clear that our phenomenological theory, utilizing the normal
state $dc$ resistivities as the input parameters for the microscopic
calculation of the dielectric response, gives a good semiquantitative
description of the c-axis optical measurements in cuprate
superconductors. Understanding the strong doping dependence of the
data in ref. \onlinecite{one} requires a more complicated
phenomenological model\cite{ten} for $t_c$ and $\gamma$ because
$\rho_c$ shows, 
in general, complex doping dependence.

Finally, we point out that a simple phenomenological Casimir-Gorter
two fluid description works remarkably well for this problem. Writing
the dynamical dielectric function $\epsilon(\omega)$ as
\begin{equation}
\epsilon(\omega) = 1 - \frac{\omega_{pn}^2}{\omega^2 + i \omega
\gamma} - \frac{\omega_{ps}^2}{\omega^2},
\label{eps}
\end{equation}
where $\omega_{pn(s)}$ is the normal (superconducting) fluid c-axis
plasma frequency in the tight binding description obtained
respectively from the normal (superconducting) carrier density, we can
calculate (Fig. \ref{fig2}) the c-axis optical reflectivity. These
results shown in Fig. \ref{fig2} look qualitatively similar to our
microscopic results shown in Fig. \ref{fig1}. (The detailed temperature
dependences in the two descriptions are somewhat different.) The
physical picture that emerges from our work is thus simple: large
level broadening due to strong scattering (large $\gamma$) in the
normal state associated with ``incoherent'' transport\cite{ten} along
the c-axis overdamps the normal state c-axis plasmon preventing the
formation of a reflectivity edge and producing fairly structureless
normal state reflectivity whereas the superconducting carriers, being
unaffected by scattering effects, give rise to a sharp reflectivity
edge associated with the well-defined Anderson-Bogoliubov c-axis
plasmon which lies in the superconducting gap due to weak interlayer
carrier hopping in the layered cuprate systems.

\section*{ACKNOWLEDGMENTS}
We thank P. W. Anderson, K. A. Moler, and especially, A. J. Millis
for helpful 
discussions. This work is supported by the US-ONR.

\begin{table}

\caption{Calculated parameters $t_c$ and $\gamma$, and the
corresponding c-axis plasma
frequencies ($\omega_{pl}$) for various layered cuprates. 
}
\end{table}

\begin{figure}

\caption{ 
Reflectivity of LSCO as a function of the frequency (a) for fixed
temperature $T=0 K$ (superconducting state) with various impurity
scattering rate $\gamma$, 
(inset shows the reflectivity as a function of the frequency
normalized by the hopping parameter $t_c$ for $T=0 K$ and $\gamma =
\Delta$); 
(b) for fixed temperature $T=1.01 T_c$ (normal state) with various
$\gamma$; (c) for fixed impurity scattering rate $\gamma=\Delta$ with
different temperatures, and (d) for various T and $\gamma$.
}
\label{fig1}
\end{figure}

\begin{figure}

\caption{ 
(a) Reflectivity of $La_{2-x}Sr_xCuO_4$ ($x=0.16$) from the two-fluid
model  for different 
temperatures $t=T/T_c$ as a function of the
frequency normalized by the hopping amplitude $t_c$. (b) The same as
in (a) for $Tl_2Ba_2CuO_6$ ($T_c = 85 K$). 
}
\label{fig2}
\end{figure}

\newpage

{\bf Das Sarma and Hwang}, \hspace{1cm} {\bf Table 1}
\vspace{4cm}

\begin{center}

\begin{tabular}{||c||c|c|c|c|c|c||} \hline
Sample  & $\rho_c$ ($T_c$) m$\Omega$-cm& $\rho_{ab}$ ($T_c$) $\mu
\Omega$-cm & $t_c$ cm$^{-1}$& $\gamma$ cm$^{-1}$& 
$\omega_{pl}$ (0 K) cm$^{-1}$& $\omega_{pl}$ (exp.) cm$^{-1}$  \\ \hline\hline
La$_{1.84}$Sr$_{0.16}$CuO$_4$  & 50 \onlinecite{la1} & 140
\onlinecite{la1} & 23 & 330 & 57 & 55 \onlinecite{one} \\ \hline 
Tl$_2$Ba$_2$CuO$_6$ & 100 \onlinecite{tl1} & 100 \onlinecite{tl1}& 12
& 300 & 30  & $\leq$ 30 \onlinecite{tl2}\\ \hline 
YBa$_2$Cu$_3$O$_6.6$   & 200 \onlinecite{y1} & 100 \onlinecite{y1}& 9
& 570 & 55 & 60 \onlinecite{y2}\\ \hline 
Bi$_2$Sr$_2$CaCu$_2$O$_8$ &$10^4$ \onlinecite{bi1} & 25
\onlinecite{bi1} & 0.3 & 100 & 5 & $\leq$ 7.5 \onlinecite{bi2}\\ \hline
\end{tabular}

\end{center}

\end{document}